
\input phyzzx
%
\catcode`\@=11 
\def\papers{\papersize\headline=\paperheadline\footline=\paperfootline}
\def\papersize{\hsize=40pc \vsize=53pc \hoffset=0pc \voffset=1pc
   \advance\hoffset by\HOFFSET \advance\voffset by\VOFFSET
   \pagebottomfiller=0pc
   \skip\footins=\bigskipamount \normalspace }
\catcode`\@=12 
\papers
%

\def\square{\kern1pt\vbox{\hrule height 1.2pt\hbox{\vrule width 1.2pt\hskip 3pt
   \vbox{\vskip 6pt}\hskip 3pt\vrule width 0.6pt}\hrule height 0.6pt}\kern1pt}

\def\bra#1{\langle #1 |}
\def\ket#1{| #1 \rangle}

\def\ov{{\overline}}
\def\A{{\cal A}}
\def\B{{\cal B}}
\def\C{{\cal C}}

\def\H{{\cal H}}

\def\L{{\cal L}}
\def\M{{\cal M}}

\def\P{{\cal P}}

\def\R{{\cal R}}
\def\V{{\cal V}}
\def\bz{{\overline z}}

\def\ov{\overline}

\def\wt{\widetilde}
\def\wh{\widehat}
\overfullrule=0pt
\baselineskip 13pt plus 1pt minus 1pt
\pubnum{MIT-CTP-2206\cr
hep-th/9305026}
\date{May 1993}
\titlepage
\title{CLOSED STRING FIELD THEORY: ~AN INTRODUCTION}
\author{Barton Zwiebach \foot{Supported in part by D.O.E.
contract DE-AC02-76ER03069.}}
\address{Center for Theoretical Physics \break
Laboratory of Nuclear Science\break
and Department of Physics\break
Massachusetts Institute of Technology\break
Cambridge, Massachusetts 02139, U.S.A.}
\abstract{In these introductory notes I explain some basic
ideas in string field theory. These include: the concept of a string field,
the issue of background independence,
the reason why minimal area metrics solve the problem of generating all Riemann
surfaces with vertices and propagators, and how Batalin-Vilkovisky structures
arise from the state spaces of conformal field theories including ghosts.
More advanced topics and recent developments are summarized.
(To appear in the proceedings of the 1992 Summer School at Les Houches.)}

\endpage

\REF\siegel{W. Siegel, `Introduction to String Field Theory' (World Scientific,
Singapore, 1988).}
\REF\zwiebachl{B. Zwiebach, `Closed string field fheory: Quantum action and
the Batalin-Vilkovisky master equation', Nucl. Phys {\bf B390} (1993) 33,
hep-th/9206084.}
\REF\saadizwiebach{M. Saadi and B. Zwiebach, `Closed string
field theory from polyhedra', Ann. Phys. {\bf 192} (1989) 213.}
\REF\kks{T. Kugo, H. Kunitomo and K. Suehiro, `Non-polynomial closed string
field theory', Phys. Lett. {\bf 226B} (1989) 48.}
\REF\kugosuehiro{T. Kugo and K. Suehiro,  `Nonpolynomial closed
string field theory: action and gauge invariance', Nucl.
Phys. {\bf B337} (1990) 434.}
\REF\kaku{M. Kaku, `Geometrical derivarion of string field theory from
first principles: closed strings and modular invariance. Phys. Rev.
{\bf D38} (1988) 3052;\hfill\break
M. Kaku and J. Lykken, `Modular Invariant closed string field theory',
Phys. Rev. {\bf D38} (1988) 3067.}
\REF\zwiebachqcs{B. Zwiebach, `Quantum closed strings from minimal
area', Mod. Phys. Lett. {\bf A5} (1990) 2753.}
\REF\zwiebachtalk{B. Zwiebach, `Recursion Relations in Closed String
Field Theory', Proceedings of the ``Strings 90'' Superstring Workshop.
Eds. R. Arnowitt, et.al. (World Scientific, 1991) pp 266-275.}
\REF\sonodazwiebach{H. Sonoda and B. Zwiebach, `Closed string field theory
loops with symmetric factorizable quadratic differentials',
Nucl. Phys. {\bf B331} (1990) 592.}
\REF\zwiebachma{B. Zwiebach,  `How covariant closed string
theory solves a minimal area problem',
Comm. Math. Phys. {\bf 136} (1991) 83 ; `Consistency of closed string
polyhedra from minimal area', Phys. Lett. {\bf B241} (1990) 343.}
\REF\ranganathan{K. Ranganathan, `A Criterion for flatness in minimal
area metrics', Comm. Math. Phys. {\bf 146} (1992) 429. }
\REF\wolfzwiebach{M. Wolf and B. Zwiebach, `The plumbing of minimal
area surfaces', IASSNS-92/11, submitted to Comm. Math. Phys. hep-th/9202062.}
\REF\strebel{K. Strebel, ``{\it Quadratic Differentials}'', Springer-Verlag
(1984)}
\REF\gromov{M. Gromov, Filling Riemannian manifolds, Jour. Diff. Geom.,
{\bf 18} (1983) 1-147.}
\REF\calabi{E. Calabi, Extremal isosystolic metrics for compact surfaces,
preprint.}
\REF\thorn{C. B. Thorn, `Perturbation theory for quantized
string fields', Nucl. Phys. B287 (1987) 61.}
\REF\thornpr{C. B. Thorn, `String field theory', Phys. Rep.
{\bf 174} (1989) 1.}
\REF\bochicchio{M. Bochicchio, `Gauge fixing for the field
theory of the bosonic string', Phys. Lett. {\bf B193} (1987) 31.}
\REF\hata{H. Hata, `BRS invariance and unitarity in closed
string field theory', Nucl. Phys. {\bf B329} (1990) 698;
`Construction of the quantum action for path-integral
quantization of string field theory', Nucl. Phys. {\bf B339} (1990) 663.}
\REF\alvarez{L. Alvarez-Gaume, C. Gomez, G. Moore and C. Vafa, Nucl.
Phys. {\bf B303} (1988) 455;\hfill\break
C. Vafa, Phys. Lett. {\bf B190} (1987) 47.}
\REF\nelson{P. Nelson, Phys. Rev. Lett. {\bf 62} (1989) 993;\hfill\break
H. S. La and P. Nelson, Nucl. Phys. {\bf B332} (1990) 83;\hfill\break
J. Distler and P. Nelson, Comm. Math. Phys. {\bf 138} (1991) 273.}
\REF\hatazwiebach{H. Hata and B. Zwiebach, `Developing the covariant
Batalin-Vilkovisky approach to string theory' MIT-CTP-2184, to appear
in Annals of Physics. hep-th/9301097.}
\REF\schwarz{A. Schwarz, `Geometry of Batalin-Vilkovisky quantization', UC
Davis preprint, hep-th/9205088, July 1992. }
\REF\witten{E. Witten, `On background independent open-string field theory',
hep-th/9208027\hfill\break
`Some computations in background independent open-string field fheory',
hep-th/9210065.}
\REF\liwitten{K. Li and E. Witten. `Role of short distance behavior in
off-shell open string field theory', IASSNS-HEP-93/7.}
\REF\shatashvili{S. Shatashvili,`Comment on the background independent
open string theory', IASSNS-HEP-93/15.}
\REF\wittenzwiebach{E. Witten and B. Zwiebach, `Algebraic structures
and differential geometry in 2D string theory',
Nucl. Phys.{\bf B377} (1992) 55. hep-th/9201056.}
\REF\everlinde{E. Verlinde, `The master equation of 2D string theory'
IASSNS-HEP-92/5, to appear in Nucl. Phys. B. hep-th/9202021.}
\REF\stasheff{J. Stasheff, `Homotopy associativity of H-spaces, II.',
Trans. Amer. Math. Soc., {\bf 108} (1963) 293; `H-Spaces from a homotopy
point of view', Lecture Notes in Mathematics
{\bf 161}, Springer Verlag, 1970;\hfill\break
See also: T. Lada and J. Stasheff, `Introduction to sh Lie algebras for
physicists', hep-th/9209099.}
\REF\sonodazw{H. Sonoda and B. Zwiebach, `Covariant closed string
theory cannot be cubic', Nucl. Phys. {\bf B336} (1990) 185.}
\REF\lianzuckerman{B. H. Lian and G. J. Zuckerman, `New perspectives on the
BRST-algebraic structure
of string theory' Yale preprint, November 1992, hep-th/9211072.}
\REF\getzler{E. Getzler, `Batalin-Vilkovisky algebras and two-dimensional
topological field theories' MIT preprint, hep-th/9212043.}
\REF\penkavaschwarz{M. Penkava and A. Schwarz, `On some algebraic structures
arising in string theory', UC Davis preprint, hep-th/9212072.}
\REF\horava{P. Horava, `Spacetime diffeomorphism and topological $w_\infty$
symmetry in two dimensional topological string theory', EFI-92-70, January
1993.}
\REF\sen{A. Sen, Nucl. Phys. {\bf B345} (1990) 551; {\bf B347} (1990) 270.}
\REF\kugozwiebach{T. Kugo and B. Zwiebach, `Target space duality
as a symmetry of string field theory',  Prog. Theor. Phys. {\bf 87}
(1992) 801, hep-th/9201040.}
\REF\rangasonoda{K. Ranganathan, H. Sonoda and B, Zwiebach, `Connections on
the state-space over conformal field theories', MIT-preprint MIT-CTP-2193,
April 1993, hep-th/9304053.}

\chapter{The Origin of the String Field}

Much of the work in string theory has been done in the
context of first quantization. This means working with
two dimensional field theories, and making use of some
interpretation in order to relate observables in the
two dimensional theory, the world-sheet theory, to observables
in the physical spacetime theory, the target space theory.
This two dimensional field theory, typically a conformal
field theory, is the two dimensional analog of one-dimensional
field theories that describe the classical
mechanics of point particles.
Most of particle physics, however, is not done in first quantization.
Our description of Yang-Mills theories, or gravitation, is
done in the context of relativistic quantum field theory, or second
quantization. String field theory aims to formulate string theory as
a spacetime field theory. It follows that string field theory is the
{\it unconventional} approach to string theory based on the idea of
extending and generalizing the {\it conventional} field theory approach
to particle physics.

One must certainly keep in mind that generalizations of the usual
particle field theory concepts are necessary when one formulates
string field theory. For example, in classical mechanics one writes
an action for the coordinates $x^\mu (\tau )$ of a point
particle, and in the passage to field theory one
defines the field $\phi (x^\mu )$, dropping the proper
time variable $\tau$. For strings the analog is to pass
from $X^\mu (\sigma , \tau )$ to {\it functional} fields
$\Phi (X^\mu (\sigma ))$. This is well known not to work,
the string field must have extra arguments that include the
ghost fields of reparametrization invariance (see [\siegel ]). Only for
such string fields one can write suitable string field
actions. Further surprises were found in writing closed
string field theory (see [\zwiebachl ]), and very likely additional ones are
awaiting us in our way to a complete formulation.
Thus, in the final formulation, string field theory may have
little in common with standard formulations of particle
field theory except for the use of fields as dynamical
variables. We can also hope, since the physics of strings
is different from that of particles, that nonperturbative
string phenomena may be extracted from string field theory
less painfully than is typically the case in particle field theory.

It is very important to emphasize that, in current approaches
to string field theory, we work
in the path integral approach. That is, we do not work with
string field operators that create or destroy strings, but
rather with string fields which are classical $c$-numbers
(and classical anticommuting numbers, for ghost fields or
fermi fields). Quantization is defined by doing path
integrals with the string field action.

Let us consider in some detail the type of first-quantized actions
used to describe string theory. This will allow us to understand the
origin of the string field. One of the simplest actions used to
describe strings in first quantization is the following

$$S \sim \int d^2\xi \, \sqrt{h} h^{ab} \partial_a X^\mu
\partial_b X^\nu \, \eta_{\mu\nu}, \eqn\simplfqa$$

\noindent
where $a,b=1,2$ are indices labeling the coordinates $\xi^a$ in the
two dimensional surface,
$h_{ab}$ is a two dimensional metric, the scalar fields
$X^\mu (\xi)$ give the embedding of the worldsheet in spacetime,
and $\eta_{\mu\nu}$ is the flat Minkowski spacetime metric.
String theory is supposed to be a theory of gravity so one
may ask, where is the graviton, or the metric tensor in the
above action ? It is not there. That is the case because this is a first
quantized action. One finds that this action can be used
in a very indirect way to describe gravitons.

There are two
dimensional actions that provide a somewhat more explicit
framework for the study of the dynamics of the target space
fields. These are the sigma model actions, and typically
read as follows

$$\eqalign{
S' \sim \int d^2\xi \,\bigl(&
\sqrt{h} h^{ab} \partial_a X^\mu \partial_b X^\nu \, G_{\mu\nu}(X)
+\epsilon^{ab} \partial_a X^\mu
\partial_b X^\nu \, B_{\mu\nu}(X)\cr
& + \sqrt{h} R^{(2)}(h) \Phi(X) +
\cdots \bigr). \cr}\eqn\lsimplfqa$$

\noindent
Here we see that the Minkowski metric $\eta_{\mu\nu}$ was replaced by the
arbitrary metric $G_{\mu\nu}(X)$ and extra interactions were
added, including in particular one parametrized by an antisymmetric
object $B_{\mu\nu}(X)$, and another by $\Phi(X)$. In a sigma model action,
the objects $G_{\mu\nu}(X)$, $B_{\mu\nu}(X)$, and $\Phi(X)$, are
prescribed nonlinear functions of the two dimensional field $X(\xi )$.
The above action is therefore that of a nonlinear sigma model.
The object $G_{\mu\nu}(X)$
should play the role of a metric tensor in spacetime, $B_{\mu\nu}(X)$
should correspond to an antisymmetric tensor in spacetime, and $\Phi(X)$
should correspond to a scalar field, a dilaton in spacetime. Now we have the
metric
tensor, but where are Einstein's equations for $G_{\mu\nu}$, or the equations
for the antisymmetric tensor $B_{\mu\nu}$, or the equations for the
dilaton field $\Phi$ ? They are not found directly from the action,
again because this is not a second quantized action. They arise,
however, from a rather remarkable condition. The condition that
the above action be conformal invariant yields Einstein's equations
for $G_{\mu\nu}$ along with equations for the other fields.
While fascinating, it seems very difficult to use this approach for
a complete formulation of string theory. The complete classical equations
for the spacetime fields involve calculations to all loops in the
two dimensional field theory. The method is also very difficult
to use for spacetime fields that are not massless (we did not include
those in the above action) since they correspond to nonrenormalizable
interactions in two dimensions.

Nevertheless we see a general pattern. Each possible two dimensional
interaction, or local operator, is accompanied by a spacetime field
which appears as a coupling ``constant'' multiplying the interaction.
If we think of the string field as a collection of spacetime fields, the
above action suggests that the string field simply encodes the
data of a two dimensional field theory. Given a particular
string field we can associate a particular two dimensional field
theory. It seems very difficult at this stage to make precise
this idea for a variety of technical and conceptual
difficulties. The way we {\it know} how to proceed goes
as follows.

If we have a conformal field theory, as is the case for the first
action we wrote above,\foot{We must include, however, the reparametrization
ghosts.} conformal invariance provides us the
complete and explicit list of all possible local operators $\Phi_s$ of
the two dimensional conformal theory. For each of these
operators we associate a spacetime field $\psi_s$. The state space
$\H_{\hbox{CFT}}$ of a conformal field is the space of states created by
letting all
possible local operators act on the vacuum. Such states are denoted
by $\ket{\Phi_s}$. For example, to the familiar graviton vertex operator
$\partial X^\mu\overline\partial X^\nu e^{ipX}$ we must
associate the graviton field $h_{\mu\nu}(p)$
$$\partial X^\mu\overline\partial X^\nu e^{ipX}\leftrightarrow h_{\mu\nu}(p).
\eqn\assvof$$

\noindent
We are then naturally led to assemble the string field as a general
vector in the state space $\H_{\hbox{CFT}}$
$$\ket{\Psi} = \sum_s \ket{\Phi_s}\, \psi_s \,\, .\eqn\setupsf$$
Here each {\it target space field} $\psi_s$ is the component
of the vector $\ket{\Psi}$ along the basis vector $\ket{\Phi_s}$.
The target space fields are in general complex numbers, and may be
Grassmann even or odd. The string field action $S(\Psi )$ is a function
from $\H_{\hbox{CFT}}$ to the real numbers.

Loosely speaking, the string field does indeed appear to encode the data of
a two dimensional field theory. If $S_c$ denotes the action
for the conformal field theory that is being used to define
the string field, then associated to the string field in
Eqn.\setupsf\ we may attempt to define the two dimensional action
$$S_c+ \sum_s \, \psi^s \cdot \int d^2 \xi\, \Phi_s(\xi ,\bar \xi),
\eqn\actdef$$
using the components of the string field (the spacetime fields)
to weight the various interactions.\foot{I am ignoring ghost insertions that
are necessary to go from the string field to the world sheet interactions.}
Again, due to technical difficulties it is difficult
to define precisely the meaning of the above expression, unless
the operators $\Phi_s$ are primary fields of dimension (1,1).
Therefore, at present, the precise formulation of closed string
field theory does not attempt to define the string action as
a function in the space of two dimensional theories, but rather
as a function in the state space of a given conformal field theory.

Therefore, in order to write a string field theory, we must first choose
a conformal field theory. A conformal field theory defines a consistent
spacetime background for string propagation. This means that we are only
able to write string field theory once we pick a background that must
correspond to a classical solution of the theory we are aiming to write.
An analogy is useful to understand the situation.
In Einstein's gravity the dynamical variable is the metric tensor
$g_{\mu\nu}(x)$ on some manifold $M$. The Einstein action, given a metric
$g_{\mu\nu}$ gives us a number, this action is therefore a function on the
space ${\cal G}$ of metrics on $M$. There are some special metrics on
${\cal G}$, the Ricci flat ones. They solve the classical Einstein's equations
and therefore define consistent spacetime backgrounds.
We can study physics around any Ricci flat background $\hat g_{\mu\nu}$ by
expanding the metric tensor as $g_{\mu\nu} = \hat g_{\mu\nu} + h_{\mu\nu}$,
where $h_{\mu\nu}$ represents a fluctuation. The gravity action becomes a
function of the field $h_{\mu\nu}$. While this action for $h_{\mu\nu}$ does
contain all of the physics of gravity, the action depends explicitly of
the background metric $\hat g_{\mu\nu}$ in a complicated way. In string
theory the analog of a given metric $g_{\mu\nu}$ is a two dimensional
field theory, and the analog of the space ${\cal G}$ of metrics  on $M$
is the space of two dimensional field theories (it is not clear how to think
of a two-dimensional field theory as a structure in some space, thus
there is no compelling analog for $M$). Corresponding to the Ricci-flat
metrics we have the conformal field theories. The string field $\ket{\Psi}$
we have discussed above
corresponds to the fluctuation field $h_{\mu\nu}$.  Indeed the string field
action has background dependence; it uses, for example, the BRST operator of
the
conformal field theory. This necessity to fix a conformal field theory to get
started writing a string field action is usually referred to as the issue
of background independence of string field theory. It is certainly the
central question facing string field theory. A background independent
string field theory would most likely be the formulation of string theory we
are looking for.

The problem of setting up a background independent string field theory
is exactly analogous as that of reconstructing Einstein's theory if
we only knew the expansion of the Einstein lagrangian around flat space.
Of course, the Einstein-Hilbert action was written before the flat space
expansion was known, but as a problem of principle, the issue of reconstructing
the gravity action from the interactions of a spin two field $h_{\mu\nu}$
has received considerable attention. Our current problem, and in this case
we do not know the answer in advance, is to find the action from which
our string field action $S(\Psi )$ arises by expansion around a classical
solution. We need to understand the geometrical meaning of the series
of self-interactions of the string field $\Psi$.

There are various ideas that are probably going to be helpful in the
near future. They will be discussed in \S4 . Some of them center around the
Batalin-Vilkovisky (BV) theory
of quantization. The string field action $S(\Psi )$ actually satisfies
a second order, nonlinear partial differential equation called the BV
master equation. A generalization of the concept of a Lie algebra, a
structure called ``homotopy Lie algebra'' seems to be at the basis of
classical closed string field theory. At the quantum level, the algebraic
structure becomes what one could call a ``quantum homotopy Lie algebra''.
Understanding
these algebraic structures seems essential. At a geometrical level it
is probably necessary, at least in the short run, to develop more of the
geometry in the space of two dimensional field theories. Here again BV
seems to be of help; it is reasonable to assume that this theory space
is actually a (super) symplectic manifold. Finally, geometry of two
dimensional Riemann surfaces, or geometry on the space of Riemann surfaces
(moduli spaces) plays a crucial role in writing string field theory.
I would expect this geometry to tie in nicely with theory space geometry
in the future formulations of string field theory.

\chapter{String Diagrams from an Extremal Problem}

If string theory is to be formulated in a natural way as a field
theory there has to exist a natural way to generate the sum over
all surfaces (necessary in the computation of any string amplitude)
with the use of a propagator and a set of vertices. For a covariant
string field theory, the vertices should be symmetric under the
exchage of the various strings. It was widely believed that no
natural way would be found to generate all Riemann surfaces once and
only once. Indeed, experience suggested that any simple choice of
vertices and propagator would either miss surfaces, produce some surfaces
more than once (typically an infinite number of times), or both.

I would like to show here how one can build
all the different Riemann surfaces efficiently by using
vertices and propagators. I will prove explicitly two basic results
and then explain why they guarantee that all Riemann surfaces can
be produced without missing any surface and without overcounting any
surface. In order to achieve this we use string diagrams. For original
references the reader may consult [\saadizwiebach--\wolfzwiebach ].

A string
diagram is the analog of a particle field theory Feynman diagram. One speaks
of the string diagram corresponding to a given Riemann surface. The
string diagram is nothing else than the Riemann surface equipped with some
extra structure. This structure can be a analytic one-form (abelian
differential), an analytic two-form (quadratic differential), or typically,
a (conformal) metric. It is the extra structure on the Riemann surface
that should tell us how the surface is built in terms of vertices and
propagators. What we have found in the last few years is that, for
covariant string field theory, a metric seems to be the right
structure to put on a surface. That metric,
moreover, is very special, it is the solution to an extremal problem.
Not only for closed string field theory these metrics are relevant.
Extremal metrics actually define the string diagrams of open string
theory and of open-closed string field theory. Thus actually {\it all} string
diagrams for
covariant string field theory are defined by extremal metrics.

Let us first explain what is a (conformal) metric on a Riemann surface (see
[\strebel ]).
A two dimensional Riemannian manifold, that is, a two dimensional
manifold with a Riemannian metric $g_{\alpha\beta}$, actually defines
a fixed Riemann surface $\R$, since a metric defines a conformal structure.
The metric can be put locally in the conformal form $dl^2 = \rho^2(dx^2+dy^2)$,
with $z=x+iy$ the local analytic coordinate, and, with $\rho (x,y)$ the
positive semidefinite Weyl factor. $\rho$ is said to be a conformal metric
(metric, for short) on the Riemann surface $\R$. In the extremal problem we
wish to consider the complex structure will be kept fixed, and we will choose
a (conformal) metric satisfying some extremal property. Varying the Riemannian
metric would mean that we are also changing the conformal structure -this
we do not want to do.
A (conformal) metric $\rho$ on a Riemann surface $\R$ defines
a length element $dl= \rho |dz|$, and an area element
$\rho^2 dx \wedge dy$. Under analytic changes of coodinates
the metric must transform as $\rho (z) |dz| = \widetilde \rho (w) |dw|$.
This transformation guarantees the invariance of the length of a curve
and the invariance of the area under analytic mappings.

In the standard minimal area problem one has a Riemann surface $\R$, and
one chooses a set $\Gamma_i$ of free\foot{Basepoint free.} nontrivial homotopy
classes of closed
curves. Let $\widehat\gamma_i$ be representative curves for the chosen
homotopy classes. Associated to each homotopy class one chooses a constant
$A_i\geq 0$. One then asks for the metric $\rho$ of least possible
area on $\R$ under the condition that for any curve $\gamma$ freely
homotopic to $\gamma_i$ ($\gamma \sim \gamma_i$) the length of $\gamma$
on the metric $\rho$ must be greater than or equal to $A_i$:
$$\int_{\gamma \sim \gamma_i}\rho\, |dz| \geq A_i \, . \eqn\admcond$$
This requirement must hold for all $i$ in the set.
A metric is called {\it admissible} if it satisfies the length
conditions \admcond . Thus the minimal area problem asks for the
admissible metric of least possible area.
A useful property of the
space of admissible metrics is that it is a convex space; if
$\rho_0$ and $\rho_1$ are admissible metrics then
$$\rho_t = (1-t) \rho_0 + t \rho_1, \eqn\convexspace$$
is an admissible metric for all $t \in [0,1]$, as follows directly from
\admcond . Let $\A (\rho )$ denote the area of $\R$ with the metric $\rho$.

\section{Two Useful Results}

The area functional $\A$ has a very nice property: it is
strictly convex. This means that for the above one-parameter
family of metrics, with $\rho_0 \not= \rho_1$, one has
$$\A (\rho_t ) < (1-t) \A (\rho_0 ) + t \A (\rho_1 ),\eqn\strictcon$$
for $t \in (0,1)$. In words, the area is lower than the linear interpolation
of the areas at the endpoints. The inequality is strict since we do not include
the endpoints $t=0$, and $t=1$, corresponding to $\rho_0$ and $\rho_1$.
It simply means that the area, as a function of $t$, is a convex function.
This relation is simple to derive, so we will do so now. By definition
$$\eqalign{\A (\rho_t ) = \int \rho_t^2 \,\hbox{dxdy} &= \int
((1-t) \rho_0 + t \rho_1)^2 \,\hbox{dxdy}\cr
&= (1-t)^2 \A(\rho_0)+ t^2 \A(\rho_1 )+
2t\,(1-t)\int \rho_0\rho_1\,\hbox{dxdy} .\cr}\eqn\bproo$$
Since $\rho_0 \not= \rho_1$ (by assumption) Schwarz's inequality implies that
$$\int \rho_0\rho_1\,\hbox{dxdy}< \Bigl( \int \rho_0^2\,\hbox{dxdy}
\int \rho_1^2\,\hbox{dxdy}\Bigr)^{1/2} = \bigl( \A (\rho_0) \A (\rho_1)
\bigr)^{1/2} .\eqn\schwarzi$$
Using this inequality for the last term in \bproo\ we find
$$\sqrt{\A (\rho_t )} < (1-t) \, \sqrt{\A (\rho_0 )} + t\, \sqrt{\A (\rho_1 )},
\eqn\sqiscon$$
which shows that $f(t) = \sqrt{\A (\rho_t)}$ is convex. But, if an everywhere
positive function is convex, its square is also convex. This implies the
desired result, namely, the area functional is convex.

An immediate consequence
is the uniqueness of the minimal area metric (if it exists).
This is proven as follows. Assume there are two different metrics
$\rho_0$ and $\rho_1$ on the surface $\R$, both admissible and
both having the (same) least possible area $\A$. It follows that
the admissible metric $\rho_t$ (for any
$t\in (0,1)$ )  must satisfy
$\A (\rho_t) < (1-t) \A + t\A = \A$, in contradiction with the assumption
that $A$ was the least possible value for the area.
The uniqueness of the minimal area metric is fundamental for our purposes,
as will be explained later.

There is another result, which is also standard, easy to derive, and
very powerful. Consider the rectangular region in the complex plane
bounded by the points $0,a,ib$, and $a+ib$, where $a$ and $b$ are
real numbers greater than zero. On this region of the complex plane
we would like to find the metric of least possible area such that any
curve starting anywhere on the left vertical segment $[0,ib\, ]$ and ending
anywhere on the right vertical segment $[a,a+ib\, ]$ should be longer than
or equal to $a$. This problem is the canonical (and simplest) minimal
area problem. We are looking for a metric $dl^2 = \rho^2 (\hbox{dx}^2
+ \hbox{dy}^2)$ with least possible area $\A (\rho ) = \int_0^b \hbox{dy}
\int_0^a \hbox{dx}\,\rho^2$. It is clear that the metric $\rho(x,y) =1$
is admissible, indeed, any curve crossing the rectangle from left to right
must be longer than or equal to the length of a strictly horizontal line
($y$ constant). The horizontal lines are precisely of length $a$. The
area of this metric is $ab$. It follows that for the metric $\hat \rho$
of least possible area we must have $\A (\hat \rho ) \leq ab$. We will
show now that actually $\rho = 1$ is the metric of least possible area.
To this end consider an admissible metric $\rho$. Since the length condition
must be satisfied for a horizontal curve with fixed $y$
$$\int_0^a \rho (x,y)\, \hbox{dx} \geq a, \quad \forall y \, ,\,\quad\to\quad
\int_0^b \hbox{dy} \int_0^a \rho (x,y)\, \hbox{dx} \geq ab \, .\eqn\ffcond$$
The Schwarz inequality
$$\int f^2 \hbox{dxdy}\int g^2 \hbox{dxdy}
\geq \Bigl( \int f\, g \, \hbox{dxdy} \Bigr)^2  ,\eqn\sineq$$
with the choice $f= \rho$, and $g=1$, gives us
$$\A (\rho ) =  \int \rho^2 \hbox{dxdy}
\geq \, {1\over ab} \, \Bigl( \int \rho\, \hbox{dxdy} \Bigr)^2 .\eqn\neem$$
All integrals here are over the rectangular region. Using \ffcond\ we
immediately see that for any admissible metric $\rho$ we have that
$\A (\rho) \geq ab$. Thus the least possible area is indeed $(ab)$. Since
that was the area for the metric $\rho(x,y) = 1$, the uniqueness of the
minimal area metric implies that the minimal area metric is indeed
$\rho (x,y ) =1$. This is what we wanted to show.

We can get a further result as a simple consequence of the above.
Identify the left edge and the right edge of the rectangle via the
relation $iy \equiv a+iy$ , for $0\leq y\leq b$. We then get an annulus, or
a cylinder.
It is clear from the above derivation that the metric of least possible
area under the condition that any closed curve, freely homotopic to
a horizontal closed curve, be longer than or equal to $a$, is still the
flat metric $\rho (x,y) = 1$.

\section{The Minimal Area Problem for Closed String Theory}

We discussed at the beginning of \S2 what is the
general minimal area problem. One chooses a set (finite or infinite)
of homotopy classes of closed curves and imposes length conditions
on all curves belonging to the chosen homotopy classes. This problem
has not yet been solved, in fact, very little is
known. The solution is known for the case when we choose a set of
homotopy classes having representative curves that can be chosen
to be nonintersecting simple closed Jordan curves (no self intersections).
Such a set of representative curves is called, in
the mathematical literature, an admissible set. For any fixed genus
the maximum number of curves in an admissible set is finite --if we
try to add additional curves we must produce intersections. If we
impose length conditions on curves homotopic to those on an admissible set
the minimal area metric is known. It is a metric that arises from a
Jenkins-Strebel quadratic differential [\strebel]. Intuitively, this
means that the surface, with the minimal area metric, is built by gluing
together flat euclidean cylinders of circumferences equal to the length
parameters $A_i$ (see equation \admcond ). If we put length conditions
on curves homotopic to representatives that must intersect, even on
two such representatives, the solution is not known.

The minimal area problem relevant for closed string field theory is
the following. For any
surface $\R$ (with or without punctures) we are interested in the
metric of least possible area under the condition that
{\it all} nontrivial closed curves on $\R$ be longer than or
equal to $2\pi$ [\zwiebachma ]. This minimal area problem, in contrast
with the related extremal problems studied earlier
[\strebel ], does not require specifying some homotopy classes of
curves on the Riemann surface on which we impose length conditions.
The length conditions are imposed on {\it all} homotopy classes. This is why
this is a problem defined on moduli space, why the problem is modular
invariant. Whenever we choose particular curves on a surface we are
introducing extra data. For example, on a torus, choosing two curves
without self intersections, and intersecting each other once, amounts to
choosing a particular representative for the torus in Teichmuller space.
Having a well defined problem which does not require extra structure
is crucial for us. Since we have put conditions on all homotopy classes
of curves, and clearly their representatives intersect, this problem does
not fall within the class of problems whose solution is known.

The results established in \S2.1 will now allow us to
prove a result that gives us a fair amount of intuition about the
minimal area problem for closed string theory.
I claim that if a surface is built by gluing
together flat cylinders of circumference $2\pi$, and moreover, no closed
curve on the resulting surface is shorter than $2\pi$, then the surface
has a metric solving the minimal area problem. The argument (which is
truly simple) goes as follows. The cylinders split the surface into
a set of rectangular domains of the type considered in the previous
section, each with $a=2\pi$, and with the vertical edges identified. Since
the core curve of each rectangle (a closed horizontal curve) is nontrivial,
it is necessary that any curve homotopic to a core curve
and contained completely on the corresponding rectangle be longer than or
equal to $2\pi$. Let us call this condition $(\alpha )$. Condition
$(\alpha )$ is necessary, although not sufficient for a metric to be
admissible. Imagine
there is an admissible metric with area lower than that of the
original metric. Then it
would have to have lower area at least on one of the rectangles. But this is
impossible since, on each rectangle, the original flat $\rho =1$ metric  has
already the least possible area under condition $(\alpha )$. This proves the
result. We therefore have a simple way to construct many metrics of minimal
area.
We glue flat cylinders watching out that no closed curve is smaller than
$2\pi$.

Let us now explain a very fundamental idea, the idea that shows the
relevance of minimal area metrics. The origin of all the difficulties
in constructing a field theory of closed strings lies on the fact that given
two Riemann surfaces it is, in general, very hard to tell whether or
not they are the same, that is, whether or not there is a conformal
map from one to the other. This means that when we construct
Riemann surfaces using vertices and propagators it is very hard to guarantee
that no Riemann surface is produced more than once. The situation is
very much improved if we put metrics on the surfaces. It is actually
easy to see if two surfaces with metrics are the same. For example, if
we build our surfaces using the cylinders discussed in the paragraph
above two metrics are the same if {\it they look the same}. That is,
they must have the same number of cylinders, and the gluing patterns
must be the same. This is the case because the cylinders determine
very special {\it saturating geodesics} on the surface, length $2\pi$
geodesics that saturate the length conditions. Two surfaces cannot have
the same metric if their patterns of saturating geodesics are not the same.
When we construct
our surfaces with metrics, our problem is making sure that for any
two such surfaces $(\R_1,\rho_1)$, and $(\R_2,\rho_2)$ the underlying
Riemann surfaces $\R_1$ and $\R_2$ are not the same (otherwise we have
overcounting). Suppose we can
tell that the metrics are not the same, what does this buy for us?
If the metrics are not the same and the Riemann surfaces are different
that is no problem\foot{Precisely speaking, in this case, the metrics
are different as Riemannian metrics on the underlying two dimensional
manifold, since as conformal metrics they are not defined on the same
Riemann surface and any comparison is meaningless.}. The problem happens
if the Riemann surfaces are the same despite the fact that the metrics
are different.
Here is where the minimal area principle helps; if we can assure that
the metrics are of minimal area, their being different guarantees that
the Riemann surfaces are different! The reason is uniqueness of the
minimal area metric on a given Riemann surface. If the two Riemann surfaces
were the same they could not
have two different metrics solving our extremal problem. As one can
imagine, different Feynman graphs in string theory correspond to
different length of cylinders and different gluing patterns. Therefore
different Feynman graphs produce different metrics, and if we guarantee
that they are all of minimal area, we are free of the problem of overcounting.
I have explained above one simple criterion to tell whether a metric is
of minimal area (the surface is built with cylinders). This criterion is
sufficient but not necessary, in fact
not all minimal area metrics are of this type.
In the next section I will sketch an argument why the rules of
sewing, which allow us to build complicated surfaces starting from
vertices and propagators, are compatible with minimal area.

A very interesting problem, an extremal problem for metrics on
Riemannian manifolds, has been investigated by Gromov [\gromov ] and
Calabi [\calabi ]. Their problem for the case of two dimensional surfaces
is a particular case of our minimal area problem. In their case they
consider a fixed two dimensional manifold $M$ and the space of
Riemannian metrics on $M$ such that all nontrivial closed curves are
longer than or equal to $2\pi$. Then they ask for the Riemannian metric
of least possible area. Since they can change the complex structure of
the surface, it seems clear that their ``extremal isosystolic'' metrics
should correspond, at every genus, to a minimal area metric on the Riemann
surface whose extremal metric has the least possible area.
They can prove existence of the
extremal isosystolic metric. Uniqueness is not clear, it could be
that a finite number of different Riemann surfaces have extremal metrics
whose area is the same, and at the same time lowest among all
the areas of the extremal metrics on all other Riemann surfaces.

\section{Why Minimal Area Metrics Work}

Let us now sketch the logic that shows that minimal area metrics
solve our problem of generating all Riemann surfaces once and only once.
Here I will not be completely self-contained, but hope to give
a reasonably clear idea of how things fit together. For complete details
the reader should consult the original papers.

A metric solving the minimal area problem is expected to
give rise to closed geodesics of length $2\pi$ that foliate
the surface completely. If there is a point through which there is no
saturating geodesic the
metric could be lowered at this point without destroying
admissability.
We can show that such curves must foliate the surface since
they cannot intersect whenever they are of the same homotopy
type. In fact, generically, two saturating geodesics can at most
intersect in one point.
For the case of Riemann spheres with punctures,
since any Jordan closed curve must cut the surface into two separate
pieces, two saturating geodesics that cross they must do so
at least in two points. Since this cannot happen,
the surface will be foliated by bands of geodesics that do not
intersect. This is precisely what happens with Jenkins-Strebel (JS)
quadratic differentials, the horizontal trajectories of the
quadratic differential are the saturating geodesics. The surface
is then foliated by bands of geodesics, the so-called ring-domains
of the quadratic differential. The horizontal trajectories
can intersect in the critical graph of the quadratic differential,
but this graph is of zero measure on the surface. Thus all minimal
area metrics on Riemann spheres (that define the classical closed
string field theory) arise from JS quadratic differentials. They
are now known explicitly and can be described in terms of polyhedra.
In higher genus Riemann
surfaces one can have saturating bands of geodesics
that cross. Thus higher genus minimal area metrics
do not always arise from JS quadratic differentials. One concrete
example showing  crossing of foliations was given
in [\wolfzwiebach ].

In a punctured Riemann surface, the minimal area metric must have
all closed curves homotopic to the punctures satisfying the length
conditions. This actually allows one to show that
a minimal area metric is isometric to a
semiinfinite cylinder of circumference $2\pi$ for some neighborhood
of each puncture. This semiinfinite cylinder must end somewhere on
the surface; let $\C_0$ denote the boundary
curve of the semiinfinite cylinder. Let $\C_l$ denote the saturating geodesic
in the cylinder a distance $l$ away from $\C_0$.
The minimal area metrics satisfy an amputation property; if we amputate
the semiinfinite cylinder along $\C_l$
for $l >0$, the resulting surface still has a minimal area metric.
This property allows us to show that the plumbing
of surfaces with minimal area metrics gives a surface with a
minimal area metric [\zwiebachqcs ,\wolfzwiebach]. The basic idea is
simple, given two surfaces to be sewn together (or a single surface
with two legs to be sewn), one first amputates the semiinfinite cylinders
corresponding to the relevant legs.  Given the amputation property the
resulting surfaces with boundaries have minimal area metrics. If we glue
together the open boundaries, and, by doing so we do not create closed
curves that are smaller than $2\pi$, (this is the reason for stubs, as
we will see), then the resulting surface inherits a minimal area metric.
This follows because any candidate metric with less area would have to
have less area in at least one of the surfaces that were glued, but this
is impossible given that the amputated surfaces have minimal area metrics.
The reader interested in the complete detailed argument should consult
[\wolfzwiebach ].

We define the height $h$ of a foliation (a band of geodesics making up
an annulus) to be the shortest distance, along the annulus, between the two
boundary components. It can be shown that a foliation
with height $h$ greater than $2\pi$ is isometric to a flat cylinder
of circumference $2\pi$ and height $h$. This suggests that on a string
diagram we can identify propagators with the cylinders that have
heights greater than or equal to $2\pi$. One can prove a cutting
theorem (closely related to amputation); if we cut open a foliation
of height greater than $2\pi$ on a metric of minimal area, the resulting
(cut) surface still has a minimal area metric.

When building the string
field theory we must choose vertices, we call $\V_{g,n}$ the vertex
that must be introduced at genus $g$ for processes with $n$
punctures. This vertex corresponds to the surfaces that are
missed by the Feynman rules using lower dimensional vertices. There is
a simple criterion to tell whether or not a surface $\R$ belongs to the
vertex:

\noindent
$\underline{\hbox{The String Vertex}\,\, \V_{g,n}}.\,$  $\R \in \V_{g,n}$ if
and only if the heights of all internal foliations are less than or equal to
$2\pi$. If $\R$ is in $\V_{g,n}$ we define the coordinate curves to be
the $\C_\pi$ curves around each puncture.

\noindent
In the above, internal foliation refers to a foliation
that does not correspond to one of the semiinfinite cylinders.
We have placed
coordinate curves leaving ``stubs'' of length $\pi$ for each
puncture. The coordinate curves define the amputated vertices, that
is the vertices as surfaces with holes ready to be joined to each
other with propagators. The short cylinder of length $\pi$ left
from each semiinfinite cylinder is necessary in order to make sure that the
plumbing procedure produces admissible metrics; if we had
no stubs the plumbing of two holes on a single surface,
with a propagator of small length, could introduce curves
shorter than $2\pi$, and the resulting surface would not inherit a  minimal
area metric. The stubs guarantee that
upon plumbing no curve shorter than $2\pi$ is generated. Then plumbing
is guaranteed to preserve the minimal area property.

\noindent
Two things must be checked. First, no surface in $\V$ must be produced by
sewing. Indeed, due to the stubs of length
$\pi$, sewing must create foliations
of height greater than to $2\pi$, and, by definition, the surface
cannot belong in $\V$.
The second point that must be checked is that all surfaces which
are not in $\V$ are produced by sewing, and produced only once.
The first part is clear because
for any surface which is not in $\V$ there is at least one foliation whose
height is greater than $2\pi$. By the cutting theorem we can cut all such
foliations open and obtain a surface(s) with a minimal area metric(s).
Restoring the semiinfinite cylinders around all boundaries one obtains
the Riemann surface(s) whose plumbing must give us the original surface.
Therefore no surface is missed. Finally, no surface could have been produced
more than once, since
different Feynman diagrams correspond to different metrics, which
by uniqueness of the minimal area metric, cannot correspond to the
same Riemann surface.

There are some important open questions left about the extremal
metrics solving the minimal area problem at higher genus. We do not
have a mathematical proof of the existance of such metrics. While I
am not concerned about the possibility that the metrics might not
exist,\foot{At higher genus a fraction of each moduli space is made of Riemann
surfaces whose minimal area metrics arise from quadratic differentials.
We also know now of explicit examples of minimal area metrics that
do not arise from quadratic differentials.} a proof of existance is likely to
be helpful in understanding
further properties of the extremal metrics. We would like to know
how the metrics look in general, and what type of curvature singularities
they have. It would also be interesting to know how to parametrize the
general metrics of minimal area. Explicit knowledge of the minimal area
metrics is likely to be useful in future calculations using string field
theory.

\chapter{Batalin-Vilkovisky Structures}

Batalin-Vilkovisky (BV) quantization is the complete solution
of the problem of quantizing a general gauge theory (having no
second class constraints). In the usual applications one starts
with a classical gauge invariant theory with an action $S_0 (\phi^i)$ defined
on a set of fields $\phi^i$. As a first step one extends the set of fields into
a larger set comprised of fields $\psi^s$, and antifields $\psi^*_i$. One then
finds a master action $S_M (\psi_i , \psi^*_i)$ which reduces to the original
action $S_0$ upon setting the antifields to zero, and that solves the classical
master equation $\{ S_M , S_M \} = 0$. The antibracket appearing on
the left hand side in the analog of the Poisson bracket between coordinates
and momenta in classical mechanics. Here it is defined using the pairing
between fields and antifields. To define a quantum theory, however, one must
make sure that the master action actually satisfies the complete master
equation $\hbar\Delta S_M + {1\over 2}\{ S_M , S_M \} = 0$. This typically
requires extra work, since $\Delta S_M$ need not vanish in general.
The final solution $S_M (\hbar )$ is the quantum master action.

BV quantization is extremely efficient in closed string field theory
because, once we have a consistent set of string vertices $\V_{g,n}$, we
can write directly the full quantum master action $S_M(\hbar )$ solving
the quantum master equation [\zwiebachqcs ]. The full spectrum of fields and
antifields
also arises naturally from the conformal field theory. In string field
theory there is no need to go through the steps listed in the above
paragraph.

In the present section I will explain in detail why the string field
can be broken naturally into fields and antifields. This happens because
in the state space $\H_{\hbox{CFT}}$ one can introduce a symplectic structure.
In understanding this point we will have to set up the kinetic term
of closed string field theory.

Batalin-Vilkovisky quantization in covariant open string field theory was
developed by Thorn [\thorn ,\thornpr ] and Bochicchio [\bochicchio ].
Hata found BV quantization relevant to the study of the unitarity of
the HIKKO (light-cone-style) closed string field theory [\hata ]

\section{Symplectic Vector Spaces}

In ordinary symplectic geometry, a symplectic vector space is a real
vector space $V$ equipped with a nondegenerate bilinear
two form $\omega$ (taking $V\otimes V$ to $R$). The antisymmetry property of
the form implies that given two
vectors $X, Y \in V$, $\omega (X, Y) = - \omega (Y , X)$. The nondegeneracy
property requires that whenever $\omega (X, Y) = 0,\, \forall Y$, this
must imply that $X=0$. Using a basis we can write explicitly
$\omega (X,Y) = \omega_{IJ}X^I Y^J$, and nondegeneracy implies that the matrix
$\omega_{IJ}$ is invertible. We let $\omega^{IJ}$ denote the inverse
matrix. A symplectic vector space must be even dimensional,
and one can always find a symplectic basis $(X_1,\cdots ,X_n \, ; Y^1,\cdots ,
Y^n)$ such that $\omega (X_i , Y^j) = \delta_i^j$, and, $\omega (X_i , X_j ) =
\omega (Y^i , Y^j ) = 0$. This is how the symplectic structure can be used
to provide a pairing between basis vectors.

For the cases of vector spaces whose vectors can be either even or odd
objects, as is the case for the conformal field theories relevant to us,
we must consider the super case. We then have a super vector space
with an odd, bilinear, nondegenerate two form $\omega$. The form being
odd means that $\omega (A,B)$, for vectors $A$ and
$B$ of definite statistics, is nonvanishing only if $A$ and $B$ have
opposite statistics. Bilinearity and
nondegeneracy have exactly the same meaning as in the
commuting case. Finally, the exchange symmetry of a (super) two form
requires that
$$\omega (A, B) = -(-)^{AB} \, \omega ( B,A) ,\eqn\exchpro$$
where $A,B$ appearing in the exponent refer to the Grassmanality of the
object (0[mod 2] for even objects, and, 1[mod 2] for odd objects). For even
vectors this
gives the expected minus sign. A (super) vector space equipped with
such (super) symplectic structure, as in the bosonic case, admits a
symplectic basis, which determines a pairing of basis vectors.
We want to exhibit explicitly this pairing for $\H_{\hbox{CFT}}$.

I want to show why \exchpro\ is compatible with the
exchange properties of the antibracket. The antibracket is defined
on a (super)symplectic manifold, that is, a manifold equipped with a
odd nondegenerate {\it closed} two form $\omega$. At every point on the
manifold the tangent space is a symplectic vector space. Given two
functions $\A$ and $\B$ on the manifold, the antibracket is defined
as follows
$$\{ \A , \B \} = \omega ( A , B), \eqn\antbder$$
where $A,B$ are the Hamiltonian vector fields associated to the functions
$\A ,\B$. Given the standard relation $i_{{}_A} \omega = -\hbox{d}\A$, between
functions and their corresponding Hamiltonian vectors, we note that they have
opposite
statistics. Therefore
$$\{ \A , \B \} = \omega ( A , B)= -(-)^{AB}\, \omega ( B , A)
= -(-)^{(\A+1)(\B+1)}\, \{ B , A \} ,\eqn\antbder$$
which is the correct exchange property of the antibracket.

\section{Ghost Conformal Field Theory}

The conformal field theories relevant for string theory are
those with total central charge zero. These conformal theories must
include the conformal field theory of the reparametrization ghosts,
having central charge $(-26)$, together with some other conformal theories
adding up to a central charge of $(+26)$. We need to know the
basics of this ever present ghost conformal field theory. We
consider the conformal field theory formulated in the $z$-plane
with $z= \exp (\tau + i\sigma )$.
We have ghost fields $c(z)$ and $ \ov c (\bz )$
of dimensions $(-1,0)$ and $(0,-1)$ respectively, and
antighost fields $b(z)$ and $\ov b (\bz)$ of dimensions
$(2,0)$ and $(0,2)$ respectively:
$$c(z) = \sum_n {c_n\over z^{n-1}},\quad
\ov {c}(\bz ) = \sum_n {\bar{c}_n\over\bar{z}^{n-1}},\eqn\first$$
$$b(z ) = \sum_n {b_n\over z^{n+2}},\quad
\ov {b}(\bz ) = \sum_n {\bar{b}_n\over \bar{z}^{n+2}}.\eqn\second$$
The stress tensor corresponding to this conformal theory is given by
$$T_g(z) = -2b(z)\cdot \partial c(z) -\partial b(z) \cdot c(z),\eqn\ghset$$
with a similar relation for the antiholomorphic piece. The basic operator
product expansion is
$$b(z)c(w) \sim {1\over z-w} . \eqn\opebc$$
The modes satisfy the anticommutation relations
$$\{b_n , c_m\} = \{ \bar{b}_n , \bar{c}_m \} = \delta_{m+n,0},\eqn\anticom$$
with all other anticommutators equal to zero.
It is convenient to define new zero modes by linear combinations
of the old ones
$$c_0^\pm={1\over 2}( c_0 \pm \bar{c}_0),
\quad b_0^\pm = b_0\pm\bar{b}_0.\eqn\newzm$$
One defines an in-vacuum $\ket{0} \in \H_{\hbox{CFT}}$, corresponding to $z=0$,
with no operator inserted there, and an out-vacuum $\bra{0} \in
\H^*_{\hbox{CFT}}$, corresponding to $z=\infty$, with no operator inserted
there.
It follows from the regularity of the conformal fields $c$ and $\bar c$ at
$z=0$, and $z=\infty$, that the oscillators $(c_{-1},\bar c_{-1}, c_0^+, c_0^-,
c_1, \bar c_1)$ do not annihilate the in-vacuum nor the out-vacuum. This
requires that the basic overlap be of the form
$$\bra{0} c_{-1}\bar{c}_{-1}c_0^+ c_0^- c_1 \bar{c}_1
\ket{0} \sim 1 \, . \eqn\begininner$$
We define the first quantized ghost number operator $G$ by
$$ G = 3+ \biggl[ {1\over 2} (c_0 b_0 - b_0c_0) +
\sum_{n=1}^\infty (c_{-n}b_n -b_{-n}c_n) + \hbox{a.h.} \biggr] .
\eqn\ghostdef$$
The reader should verify that $G\ket{0} = 0$. This operator satisfies
the following commutation relations
$$[G,c_n ] = c_n, \quad [G, b_n] = -b_n, \quad [G, Q] = Q, \eqn\ghostcomm$$
and analogous relations for the antiholomorphic objects.

\section{Symplectic Structure on ${\H'}_{\hbox{cft}}$}

Once the ghost conformal field theory is combined with a matter
conformal field theory, the total stress tensor is the sum of the matter stress
$(T_m(z),\ov T_m(\ov z))$, and the ghost stress tensor
$(T_g(z),\ov T_g(\ov z))$. It has central charge zero
and dimension two
$$T(z) = \sum_n {L_n \over z^{n+2}}, \quad
\ov T(z) = \sum_n {\ov L_n \over \ov z^{n+2}}.\eqn\oevo$$
The BRST operator of the combined conformal theory is given by
$$Q = \int {dz\over 2\pi i} c(z) \bigl( T_m(z)+{1\over 2}T_g(z)\bigr)
+\int {d\ov z\over 2\pi i} \ov c(\ov z) \bigl( \ov T_m(\ov z)+ {1\over 2}
\ov T_g(\ov z)\bigr),\eqn\brstdef$$
where the operators in the integrand are normal ordered.  We can verify that
$$\{ Q , b(z) \} = T_m(z)+T_g(z) = T(z),\quad
\{ Q , \ov b(z) \} = \ov T_m(\ov z)+\ov T_g(\ov z) = \ov T(\ov z).
\eqn\frbb$$

\subsection{Reflector State}
Let $\ket{\Phi_i} \in \H_{\hbox{CFT}}$ be a basis for states, and
$\bra{\Phi^i} \in \H^*_{\hbox{CFT}}$ be the dual basis:
$\bra{\Phi^j}\Phi_i\rangle =
\delta_i^j$. In any conformal theory there is an association of surfaces
to states. Consider the two punctured sphere with uniformizing coordinate
$z$, with a puncture at $z=0$, and local coordinate $z_1=z$ at that point,
and a puncture at $z=\infty$, with local coordinate $z_2 = 1/z$ at that
point. The state $\bra{R_{12}} \in \H^*_{\hbox{CFT}}\otimes \H^*_{\hbox{CFT}}$
corresponding to this surface is defined via the relation
$$\bra{R_{12}}\Phi_i\rangle_{(1)}\ket{\Phi_j}_{(2)} \equiv \langle
\Phi_i\, (z_1=0)\, \Phi_j\, (z_2=0)\rangle \equiv G_{ij},\eqn\metrdef$$
namely, the components of the state $\bra{R_{12}}$ are given by correlators
on the above two punctured sphere. The local coordinates we have chosen
are necessary to be able to define the correlators of local operators
which are not dimension zero primaries. The metric $G_{ij}$ must be
nondegenerate.
We also have the useful relations
$$\eqalign{
{}&\bra{R_{12}} (c_n^{(1)} + c_{-n}^{(2)} ) = 0,\cr
{}&\bra{R_{12}} (b_{n}^{(1)} - b_{-n}^{(2)}) = 0,\cr
{}&\bra{R_{12}} (Q^{(1)} + Q^{(2)}) = 0, \cr
{}&\bra{R_{12}} (G^{(1)}  + G^{(2)} -6) = 0.
\cr}\eqn\qonref$$
The first two identities, which hold for all $n$, can be obtained by
expressing the oscillators in terms of contour integrals of the ghosts
(or antighost) conformal fields, and using the definition \metrdef . The
last two identities also follow from contour deformation, both the
BRST charge and the ghost charge are contour integrals of holomorphic
currents. The last identity also follows directly from the first
two ones, and Eqn.\ghostdef ; it implies that a nonvanishing
overlap $\bra{R_{12}}A\rangle_{(1)}
\ket{B}_{(2)} \not= 0$ requires that the ghost numbers of $A$ and $B$
add up to six: $G(A) + G (B) = 6$.

\subsection{A Kinetic Term for Closed String Fields}
There is no unique symplectic structure that can be introduced
in $\H_{\hbox{CFT}}$. We need to understand which is the physically
relevant one. The answer emerges clearly, as I will show now, from
an analysis of equations of motion for closed string fields.

It was long felt that a satisfactory linearized closed string field equation
would be
$Q \ket{\Psi} = 0$, possibly supplemented with a ghost number constraint
for the string field. This would have been the analog of the open string
linearized field equation. It is very clear now that this could not
be the correct answer. It is interesting that the difficulty and its
resolution can be understood by attempting to write a kinetic term
for string fields that would give $Q\ket{\Psi}=0$ as an equation of motion.
Due to the nondegeneracy of the reflector an obvious choice for
kinetic term would seem to be
$$S_0^2\, \sim \,\bra{R_{12}} \Psi \rangle_{(1)} \, Q^{(2)}\,
\ket{\Psi}_{(2)}\,\quad ?\eqn\nonum$$
The problem now is ghost number. Since the vacuum was assigned
ghost number zero, and we only have operators of integer ghost number,
the ghost number of the string field should be an interger. Moreover,
the ghost numbers of $\ket{\Psi}$ and that of $Q\ket{\Psi}$ must add up
to six. This is clearly impossible, as it would require a string field
of ghost number $5/2$. Therefore the above candidate action does not
work. The fact that closed string vertex operators are usually of
the form $c(z) \ov c (\ov z) V(z,\ov z)$, with $V$ a matter operator,
suggests that the string field ought to be of ghost number $+2$. In
that case the kinetic operator $Q$ of ghost number $+1$ should be
replaced by an operator of ghost number $+2$.

A clue emerges when we note that we can restrict ourselves
to work with the subspace of the state space consisting of states that are
annihilated by $L_0 -\ov L_0$. We do not lose anything because any physical
state (a state annihilated by $Q$) which is not annihilated by $L_0 -\ov L_0$
is actually trivial (can be written as $Q$ acting on something). Indeed,
let $h-\ov h$ denote the eigenvalue of $L_0-\ov L_0$ on $\ket{\Psi}$, we then
have
$$\ket{\Psi} = (L_0-\ov L_0) \cdot {1\over h-\ov h} \,\ket{\Psi}=
\{ Q, b_0-\ov b_0 \}  \cdot {1\over h-\ov h}\, \ket{\Psi}
=  Q \cdot { b_0-\ov b_0 \over h-\ov h}\, \ket{\Psi}, \eqn\trivin$$
showing that indeed, such states are trivial. As a subsidiary
condition $L_0-\ov L_0 =0$ is not a strong condition, even with this
condition the string field still can be off the mass shell. The same
would not be true for the condition $L_0 + \ov L_0 =0$,
such a condition would impose a familiar field equation.
Our ability to impose an $L_0-\ov L_0=0$ condition suggests that we could
impose a further condition on the off shell string field; we could demand that
$(b_0 - \ov b_0) \ket{\Psi} = 0$. This condition cannot be justified
as we justified the $(L_0 -\ov L_0) = 0$ condition. We {\it cannot} prove that
a physical state which is not annihilated by $b_0-\ov b_0$ must be
trivial. Therefore this condition has a nontrivial effect on the definition
of physical states.
We may reasonably expect not to get in trouble
since on physical states $Q\ket{\Psi} = 0$, the condition
$(b_0 -\ov b_0 ) \ket{\Psi} =0$ does not lead to further constraints, as
we have already required that $L_0 -\ov L_0 = \{ Q, b_0-\ov b_0 \} $ annihilate
all states. The cohomology of $Q$ on the space of states annihilated
by $b_0-\ov b_0$ is called semirelative cohomology.

All of the pieces of the puzzle are now in place. If we impose
a $b_0^-(=b_0-\ov b_0)$ condition on all states, the overlap
$\bra{R_{12}} A\rangle \ket {B}$ would always vanish. This is
not hard to see. A state annihilated by $b_0^-$ can always be written
as $b_0^-$ acting on something ($b_0^- \ket{\alpha} = 0 \,\to
\ket{\alpha} = b_0^- c_0^- \ket{\alpha}$ since no state can be simultaneously
annihilated by $b_0^-$ and $c_0^-$). Since $b_0^-$ acting on the reflector
gives us another $b_0^-$ (Eqn.\qonref ) the relation $b_0^-b_0^- =0$
implies that we always get zero.
The solution is clear, a nondegenerate inner product must include
a $c_0^-$. We are therefore led to define
$$\langle A , B \rangle \equiv \bra{R_{12}}~A\rangle_{(1)} \, c_0^{-(2)}
\, \ket{B}_{(2)}.\eqn\inpdef$$
This bilinear inner product has the following exchange property
$$\langle A , B \rangle = (-)^{(A+1)(B+1)} \langle B , A \rangle,\eqn\skd$$
which follows from the symmetry of the reflector $\bra{R_{12}}$ under the
exchange of state spaces (see [\zwiebachl ] for details).
Restricting ourselves to ${\H'}_{\hbox{CFT}}$, the subspace of
$\H_{\hbox{CFT}}$ where all states are annihilated by $b_0^-$ and
$L_0^-$, we have that
\noindent
\subsection{Claim:~} $\langle \,\, , \,\, \rangle$ is nondegenerate on
${\H'}_{\hbox{CFT}}$.
Here is a proof. Assume $\langle A , B \rangle = 0$ , for all $\ket{A}
\in {\H'}_{\hbox{CFT}}$. It follows from
$$\langle A , B \rangle = \bra{R_{12}}~A\rangle_{(1)} \, c_0^{-(2)}
\, \ket{B}_{(2)}=0, \eqn\vanisheq$$
that the inner product actually vanishes for all $\ket{A}\in
{\H}_{\hbox{CFT}}$.
This is the case because any state that is not annihilated by $b_0^-$ can
be written as $c_0^-$ acting on a state. The $c_0^-$ acting on the
reflector gives us another $c_0^-$ which kills the $c_0^-\ket{B}$ state.
The nondegeneracy of the metric arising from $\bra{R_{12}}$ therefore
implies that $c_0^-\ket{B} =0$. Multiplying by $b_0^-$ we find that
$\ket{B}=0$, thus establishing the claim.

Another basic property of this inner product is that, on ${\H'}_{\hbox{CFT}}$
the BRST operator satisfies
$$\langle QA , B \rangle = (-)^A \langle A , QB \rangle . \eqn\brstinp$$
This is readily proven:
$$\eqalign{\langle QA , B \rangle &=
\bra{R_{12}}Q^{(1)}\ket{ A}_{(1)} \,  c_0^{-(2)} \ket{B}_{(2)}\cr
&= (-)^{A+1} \bra{R_{12}}A\rangle_{(1)} \, Q^{(2)}c_0^{-(2)} \ket{B}_{(2)}\cr
&= (-)^{A+1} \bra{R_{12}}A\rangle_{(1)}(b_0^-c_0^-+c_0^-b_0^-)^{(2)}
 \, Q^{(2)}c_0^{-(2)} \ket{B}_{(2)}\cr
&= (-)^{A+1} \bra{R_{12}}A\rangle_{(1)}\, c_0^{-(2)}b_0^{-(2)}
 \, Q^{(2)}c_0^{-(2)} \ket{B}_{(2)},\cr}\eqn\gthere$$
where we introduced $1=\{b_0^-,c_0^-\} $, and in the next step the
$b_0^-c_0^-$ term was seen to vanish upon taking $b_0^-$ into the reflector.
Since the anticommutator of $b_0^-$ with $Q$ gives $L_0^-$, which
kills $c_0^-\ket{B}$, we find
$$\eqalign{\langle QA , B \rangle
&=  (-)^A \bra{R_{12}}A\rangle_{(1)}c_0^{-(2)}
 \, Q^{(2)}b_0^{-(2)}c_0^{-(2)} \ket{B}_{(2)},\cr
&=  (-)^A \bra{R_{12}}A\rangle_{(1)}c_0^{-(2)}
 \, Q^{(2)} \ket{B}_{(2)},\cr
&=(-)^A \langle A , QB \rangle , \cr} \eqn\brstxnp$$
as we wanted to show.

All the above makes it clear that the correct kinetic term
for closed string field theory is
$$S_0^2= {1\over 2} \, \langle  \, \Psi\, , \, Q \Psi \, \rangle  \,\, ,\,\,
\quad \ket{\Psi} \in {\H'}_{\hbox{CFT}}.\eqn\kincsft$$
The equation of motion, which follows upon variation, use of the
nondegeneracy of the inner product, and Eqn.\brstinp , is $Q\ket{\Psi} =0$.
This action is gauge invariant under $\delta \ket{\Psi} = Q \ket{\Lambda}$,
with $\ket{\Lambda} \in {\H'}_{\hbox{CFT}}$. For the classical string
field theory one must restrict the sum over states in $\ket{\Psi}$ to
states of ghost number $(+2)$. It turns out that the kinetic term of the
master action is given by the same formula, the only difference being that
the string field satisfies no ghost number condition.

\subsection{Symplectic Structure in ${\H'}_{\hbox{CFT}}$.}
The physically relevant bilinear nondegenerate two-form we must choose is
therefore
$$\omega (A, B) \equiv (-)^A \, \langle\, A \, , \, B \, \rangle
.\eqn\symhcft$$
Indeed this is a bilinear, nondegenerate pairing. The sign factor in front
was introduced to get the correct exchange property
$$\eqalign{
\omega (A, B) &= (-)^A \, \langle\, A \, , \, B \, \rangle \cr
&= (-)^{A+(A+1)(B+1)} \langle\, B \, , \, A \rangle \cr
&= (-)^{A+(A+1)(B+1)+B }\, \, \omega ( B \, , \, A \, )  \cr
&= -(-)^{AB}\, \, \omega ( B \, , \, A \, ) ,\cr}\eqn\symhcft$$
as we wanted to show (Eqn.\exchpro )

\subsection{BV structure in Spacetime.}
We can now use the pairing of states in ${\H'}_{\hbox{CFT}}$ to pair
up the spacetime fields, which are nothing else but the expansion
coefficients of the string field $\ket{\Psi}$. In all generality
we may choose a basis where we pair up the states
$$\Bigl\{\,  \ket{\Phi_1}, \ket{\Phi_2}, \cdots \, \Bigr\} \, \,\,
; \, \,\, \Bigl\{\,
\ket{\wt \Phi^1},\ket{\wt \Phi^2}\cdots  \,\, \Bigr\} \eqn\pairu$$
with the condition
$$\omega \, ( \, \ket{\Phi_i} \, , \, \ket{\wt \Phi^j} \, ) = \delta _i^j .
\eqn\fgws$$
Let us denote the first group of states by $\ket{\Phi_s}$. It is
convenient to have those fields be of ghost number less than or
equal to $+2$. It then follows that the states $\ket{\wt\Phi^s}$ are
all of ghost number greater than or equal to $+3$ (with our inner
product with $c_0^-$, the nonvanishing condition demands that the ghost
numbers should add up to five). Then the expansion of the string field
reads
$$ \ket{\Psi} = \sum_s \Bigl( \, \ket{\Phi_s} \, \psi^s + \ket{\wt\Phi^s}
\, \psi^*_s\, \Bigr) .\eqn\strfld$$
The spacetime fields $\psi^s$, and spacetime antifields $\psi_s^*$ are paired.
This is how spacetime fields
and antifields arise. The string field is defined to be Grassmann even.
Since the statistics of $\ket{\Phi_s}$ and $\ket{\wt\Phi^s}$ are opposite,
the statistics of $\psi^s$ and that of $\psi^*_s$ must also be opposite.
We define the spacetime ghost number $g^t$ of a spacetime field to be equal
to $2-G$, with $G$ the ghost number of the corresponding
first quantized state. Then, we readily find that
$g^t (\psi^s) + g^t (\psi^*_s) =
4- (G_s + (5-G_s)) = -1$. These are standard properties of the BV
pairing. The reader may wish to verify that one can define explicitly
the tilde states as $\ket{\wt\Phi^j} = b_0^-\, G^{ji}\,\ket{\Phi_i}$.

In some sense this is the beginning of all the algebraic work in setting up
closed string field theory. I have tried in the above to be very
explicit about all of the basic and fundamental points. The discussion
of the construction of the complete string field action still requires
further work. In the next section I will give a brief discussion of
some of the remaining issues and of recent developments.

\chapter{Recent Developments}

We discussed in \S2 and \S3 how an extremal problem in Riemann surface
theory allows us to find string vertices, and how we go about setting
up the Batalin-Vilkovisky field-antifield structure of string field
theory. In order to write the string field theory explicitly one needs
to develop the differential geometry on an extended space
$\wh\P_{g,n}$ fibered over the space $\M_{g,n}$ of Riemann surfaces
with $n$ punctures and genus $g$ (for details consult
[\alvarez ,\nelson ,\zwiebachl ,\hatazwiebach ]). The fiber over a surface $\R$
corresponds to all possible choices of local coordinates, up to a phase, at
every puncture of $\R$. The string vertex $\V_{g,n}$ is properly thought as
a subspace of $\wh\P_{g,n}$. One can define differential forms
$\Omega^{(k)}_{\bf \Psi}$ of degree $k$ on
$\wh\P_{g,n}$. The forms are labeled by ${\bf \Psi}$, which denotes the $n$
states in ${\H'}_{\hbox{CFT}}$
to be inserted on the punctures of the
surface. These forms satisfy very nice relations that tie up the differential
geometry of $\wh\P_{g,n}$ and the algebraic structures in the conformal theory.
For example, the role of the exterior derivative $d$ is played by the BRST
operator: $d\Omega^{(k)}_{\bf \Psi} \sim \Omega^{(k+1)}_{Q {\bf \Psi}}$.
The Lie derivative along some vector field $U$ in $\wh\P_{g,n}$ is represented
by a stress energy  insertion: $\L_{{}_U} \Omega_{\bf \Psi} = \Omega_{\bf
T(u) \Psi}$, with ${\bf u}$ a `Schiffer' vector on the surface representing
the vector $U$. Finally, the role of the contraction operator on forms is
played by an antighost insertion: $i_{{}_U} \Omega^{(k+1)}_{\bf \Psi} =
\Omega^{(k)}_{\bf b(u) \Psi}$. In addition to providing conceptual
understanding, such relations simplify the verification that the string
action satisfies the master equation.

The understanding of the geometry of BV quantization has improved considerably
thanks to the work of A. Schwarz [\schwarz ]. His work allows a covariant
description, in the sense of symplectic geometry, of the BV formalism.
The BV master equation, as originally formulated reads
$$\hbar \Delta S +{1\over 2} \{ S,S\} =0 , \eqn\bveqnshort$$
where the antibracket $\{ \cdot , \cdot \}$ and the $\Delta$
operator are given by
$$\{ G , H \} =  {\partial_r G \over \partial \psi^s}
{\partial_l H \over \partial \psi_s^*}-
{\partial_r G \over \partial \psi_s^*}
{\partial_l H \over \partial \psi^s}, \quad
\Delta =(-)^{\phi_s}{\partial_l  \over \partial \psi^s}
{\partial_l \over \partial \psi_s^*}.\eqn\defbracket$$
While the antibracket has a covariant expression using the symplectic
two form of the manifold: $\{ G, H\} \sim  \partial_I G \, \omega^{IJ}
\partial_J H$, the second order differential operator
$\Delta$ is manifestly noncovariant.
The solution [\schwarz ] was to interpret $\Delta$ as an operator which
acting on functions gave the divergence of the corresponding Hamiltonian
vector field. In order to define a divergence, however, one must introduce
a volume element $d\mu = \rho \prod_I dz^I$ on the symplectic
manifold. Consequently the delta operator
$\Delta_\rho$ acquires a $\rho$ dependence. The main result of [\schwarz ]
is that $\rho$ must be chosen so that $\Delta_\rho^2 =0$. If this is
satisfied one can prove the existence of a `Darboux-Schwarz' system of
coordinates where the symplectic form $\omega$ becomes a constant and
$\rho =1$. Such system of coordinates seems necessary to carry out the
standard BV argument for the gauge independence of the observables.
In summary, a BV manifold is the object $(M,\omega ,d\mu )$, with $M$
a supermanifold, $\omega$ an  odd, symplectic, closed two-form, and $\mu$ a
volume  element that leads to a nilpotent operator $\Delta$.

As we discussed in \S1 there are indications that string field theory
may be thought as a function on the space of two dimensional field theories.
Nobody has been able to understand this space concretely. I believe it is
likely that string field theory will be eventually formulated on some abstract
space closely related to a theory space, or to theory space with some extra
conditions.
As we have seen the relevant state space of conformal theories including
ghosts have a symplectic structure. It seems clear that this state space
corresponds to the tangent space to `theory space'.   This suggests that
theory space may be a symplectic manifold. This idea was advocated by
Witten [\witten ] who also proposed a concrete construction for the
case of open string field theory. In this case, theory space is the space
of one dimensional lagrangians defined on the boundary of a disk. Rather
than finding an action $S$ (which would be a function in this theory space)
satisfying the master equation, one looks for
the corresponding Hamiltonian vector field $V_S$ . This vector must be odd and
should satisfy the condition $\{ V_S , V_S\} =0$ where the bracket is the
graded
Lie bracket. This equation, which can also be written as $V^2_S =0$, must hold
everywhere, as it guarantees that the master equation is satisfied everywhere.
At the points where $V_S=0$, we have conformal field theories.
The setup is very attractive, although it seems likely that the explicit
construction needs further work [\liwitten ,\shatashvili ]. In order to
formulate a complete quantum theory we need to obtain the full master
equation. This would mean that theory space should be a BV manifold in the
full sense, that is, an object $(M,\omega ,d\mu)$, with $d\mu$ a consistent
volume element. This volume element has not been found yet. Nevertheless
for a complete formulation the equation we must solve is not $V^2_S=0$ but
rather $V^2_S = -\hbar\, V_{\hbox{div}V_S}$ [\hatazwiebach ].

Hata and the author [\hatazwiebach ] have found the BV approach useful to
understand the
significance of the freedom to choose different sets of string vertices
$\V_{g,n}$ in order
to reproduce the sum over surfaces. More concretely, one wished to know
what is the relation between string field theories that use different
ways of breaking up the sum over surfaces. The answer is simple: two such
string field master
actions differ by terms induced by a field-antifield transformation
canonical with respect to the antibracket. It is possible
to use the differential geometry language described above in order to write
explicitly the generator of the canonical transformation in terms of the
vector field $U$ in $\wh\P_{g,n}$
generating the variation of the string vertices $\V_{g,n}$. Using the
covariant description of gauge fixing one can then show that two theories
using different decompositions of moduli space yield the
same gauge fixed action upon use of different gauge fixing conditions.

During the last year the algebraic structures that arise in closed string
field theory have been brought into the open [\wittenzwiebach  ,
\everlinde ,\zwiebachl].  The first object to emerge was a homotopy Lie
algebra.
In some sense homotopy Lie algebras (and their more familiar relatives,
the homotopy associative algebras [\stasheff ]) are the answer to a
longstanding
question. We have long suspected that string theory on-shell amplitudes
have a group theoretical meaning, indeed, special three point functions have
been related to structure constants of Lie algebras. The three point
amplitudes,
together will all the higher $n$-point functions (at the classical level)
define the structure constants of a homotopy Lie algebra~! This structure
can be readily explained. Suppose we have a graded Lie algebra
$\{ T_a , T_b \} = f_{ab}^{~c }\, T_c$. Introduce for each generator
$T_a$ an object $\eta^a$ of opposite statistics. Then consider the
following anticommuting vector field
$$V = \bigl( f_{a_1a_2}^{~~~~b} \,\eta^{a_1}\eta^{a_2} +
 f_{a_1a_2a_3}^{~~~~~~b} \,\eta^{a_1}\eta^{a_2}\eta^{a_3} + \cdots \, \bigr)\,
{\partial\over \partial \eta^b} .\eqn\hlathree$$
where the $f$'s with more than three indices are the higher structure constants
of the algebra. The condition $V^2=0$ is then imposed. This condition
gives, at lowest order, the Jacobi identity for the original structure
constants of the graded Lie algebra. At higher orders it gives an infinite
set of quadratic constraints on the structure constants. If the condition
$V^2=0$ is satisfied, we have a homotopy Lie algebra.
The physical interpretation arises when
we show that, for string theory, the structure constants are nothing else
than on-shell
scattering amplitudes, and the quadratic relations are the Ward identities
of the theory. In particular, if we label the physical states (BRST
semirelative cohomology classes) as $\ket{\Phi_a}$, the structure constant
$f_{a_1a_2\cdots  a_n}^{~~~~~~~~b}$ is simply given by the correlation
function $\langle\langle \Phi_{a_1} \Phi_{a_2} \cdots \Phi_{a_n}\wt\Phi^b
\rangle\rangle$ where the double bracket indicates that one integrates over
the positions of the punctures (this is a string amplitude). While this
homotopy Lie algebra exists on the cohomology of $Q$ in ${\H'}_{\hbox{CFT}}$,
there is a homotopy Lie algebra defined on the full state space
${\H'}_{\hbox{CFT}}$.
This is the homotopy Lie algebra that underlies classical closed string
field theory, and it corresponds to a simple modification of the previous
equation
$$V = \bigl(\, f_{a_1}^{~~b} \,\eta^{a_1}+
 f_{a_1a_2}^{~~~~b} \,\eta^{a_1}\eta^{a_2} +
 f_{a_1a_2a_3}^{~~~~~~b} \,\eta^{a_1}\eta^{a_2}\eta^{a_3}\, + \cdots\, \bigr)
\,
{\partial\over \partial \eta^b}  .\eqn\hlatwo$$
Here we have added a structure constant with two indices that can be
thought as the matrix elements of some operator: $f_{a_1}^{~~b} =
(Q)_{a_1}^{~~b}$. This time $V^2=0$ requires, as the lowest
consistency condition, that $Q$ be a nilpotent matrix. The next
consistency condition demands that $Q$
be a derivation of a product defined by the $f_{a_1a_2}^{~~~b}$'s.
The third consistency condition is quite amusing. It expresses the fact
that structure constants $f_{a_1a_2}^{~~~b}$ need
not satisfy anymore the Jacobi identity; the Jacobi identity must only
be satisfied weakly, that is, it differs from zero by terms having to
to with $Q$ acting on the structure constants representing four point
correlators. Therefore the $f_{a_1a_2}^{~~~b}$'s of this homotopy Lie algebra
do not define a Lie algebra. The reader may have guessed what is the
physical relevance of this algebra. This time the index $a$ labels a
general state $\ket{\Phi_a}$ in ${\H'}_{\hbox{CFT}}$. $Q$ is nothing
else but the BRST operator, and the higher structure constant
$f_{a_1a_2\cdots  a_n}^{~~~~~~~~b}$ is given by the off shell correlation
$\langle \Phi_{a_1} \Phi_{a_2} \cdots \Phi_{a_n}\wt\Phi^b
\rangle$ integrated over the set $\V_{0,n+1}$ defining the string vertex
coupling $n+1$ strings at genus zero.  Thus the off-shell amplitudes
of string field theory, together with the BRST operator form a homotopy
Lie algebra. I would like to emphasize that closed string field theory
seems to be necessarily nonpolynomial. It was shown by Sonoda and the
author [\sonodazw ], under very general conditions, that no cubic interaction
alone
could give a consistent four point amplitude. This actually means that
for covariant closed string field theory there is no clever choice of
$f_{a_1a_2}^{~~~b}$'s that satisfies a strict Jacobi identity. It would
seem that there is no Lie algebra underlying closed string field theory,
just a homotopy Lie algebra. I believe this point should be investigated
more closely.
Finally, consider the most general homotopy Lie algebra:
$$V = \bigl(\, f^{~b} \,+
 f_{a_1}^{~~b} \,\eta^{a_1} +
 f_{a_1a_2}^{~~~b} \,\eta^{a_1}\eta^{a_2}+ \cdots \, \bigr)\,
{\partial\over \partial \eta^b} .\eqn\hlaone$$
If we view the lowest structure constant $f^b$ as a vector $F$, and
$f_{a_1}^{~~b}$ as a matrix $Q$, this time the lowest identities
following from $V^2=0$ can be written as
$ QF = 0$, and ${Q}^2 B + F \star B  = 0$, with $B$ arbitrary, and
where $\star$ denotes the product defined with the structure constants
$f_{a_1a_2}^{~~~b}$. The reader may note that we have lost the nilpotency
of $Q$! This is the algebraic setup necessary for background independence.
Indeed, only when $f^b=0$ we recover the BRST operator, and therefore,
a CFT. This corresponds to a zero of the vector $V$ at $\eta^a=0$.
We recall that $V=0$ is indeed the equation of motion in the setup
of [\witten ]. The homotopy Lie algebra of \hlaone\ can be constructed
indirectly (and in a background dependent way) by shifting closed string
field theory with a string field which is not a classical solution
[\zwiebachl].

If we wish to consider the quantum theory this all gets generalized.
One can write a master equation for the on-shell action [\everlinde ]
and this structure can be seen to arise from the master equation
for the complete off shell action [\zwiebachl]. If we wish to use the
description using an anticommuting vector field, then,
following [\hatazwiebach ], the equation $V^2=0$ must be turned into
$V^2 = -\hbar\, v_{\hbox{div}V}$ (with $v_f$ denoting the Hamiltonian
vector corresponding to the function $f$). Quantum closed string field
theory defines a concrete realization of this algebra (constructed
in a Darboux-Schwarz frame of fields and antifields, with a density $\rho=1$).
This structure may be called a `quantum homotopy Lie algebra'. The
more desirable name of BV algebra is now being used by mathematicians
to describe a differential graded commutative algebra with
a nilpotent second order operator $\Delta$
[\lianzuckerman ,\getzler ,\penkavaschwarz ,\horava ]. The antibracket can be
defined as the failure of $\Delta $ to be a derivation.

Back to theory space and background independence. The problem of background
independence in the simplest setting is that of proving that closed
string field theories formulated around nearby conformal field theories are
actually equivalent. This is not so simple to prove, but has been achieved to
a large degree through the work of A. Sen [\sen ]. I believe a better
geometrical understanding of his result is very much needed to appreciate
well the issues involved in background independence. A host of problems
arise when one realizes that in comparing string field theories formulated
on state spaces $\H$ and $\H'$ corresponding to different conformal field
theories many natural identifications of the two spaces become meaningless
as soon as the theories are not infinitesimally away. Some of these
difficulties, whose origin lies in the fact that the state spaces are
infinite dimensional, were found in [\kugozwiebach ]. Together with
K. Ranganathan and H. Sonoda we have investigated geometry of the vector
bundle whose base manifold is a space of conformal field theories and
whose vector space at each point is the infinite dimensional state space
of the theory [\rangasonoda ]. We gave a characterization of the connections
that can be introduced in this vector bundle, and studied special connections
that allow the construction of a conformal theory using the state space of
another theory a finite distance away. I would expect connections to be
a useful ingredient in future analysis of background independence.

As we have reviewed in this section, from the algebraic viewpoint
homotopy Lie algebras and their generalizations seem to play a prominent
role in string field theory. From the geometrical viewpoint we have
seen the role of geometry on moduli spaces of Riemann surfaces,
geometry of BV quantization, and theory space geometry. I look
forward to see the fascinating relations that are likely to be
uncovered between the different geometries. Such developments
should pave the way to a complete formulation of string theory.

\ack
I am grateful to Bernard Julia for his invitation to lecture at Les Houches.

\singlespace
\refout
\end